\documentclass[a4paper,11pt]{article}
\pdfoutput=1 

\usepackage{jheparxiv}

\usepackage[T1]{fontenc} 
\usepackage{tikzit}
\tikzstyle{new style 0}=[fill=black, draw=black, shape=circle, radius=0.1]
\usetikzlibrary{shapes.geometric, arrows}

\usepackage{tensor}
\usepackage{amsmath}
\usepackage{amssymb}
\usepackage{mathrsfs}
\usepackage{mathtools}
\usepackage{bbm}
\usepackage{graphicx}
\usepackage{stmaryrd}
\usepackage{twistor}
\usepackage{xcolor}

\newcommand{\eps}{\epsilon}

\renewcommand{\sc}{\mathsf{c}}

\newcommand{\AdS}{\mathrm{AdS}}
\newcommand{\cI}{\mathcal{I}}
\newcommand{\LH}{L\mathfrak{ham}(\mathbb{C}^2)}
\newcommand{\LHL}{\mathfrak{ham}_\Lambda(\mathbb{C}^2\times \C^*)}
\makeatletter
\newsavebox{\@brx}
\newcommand{\llangle}[1][]{\savebox{\@brx}{\(\m@th{#1\langle}\)}%
  \mathopen{\copy\@brx\kern-0.5\wd\@brx\usebox{\@brx}}}
\newcommand{\rrangle}[1][]{\savebox{\@brx}{\(\m@th{#1\rangle}\)}%
  \mathclose{\copy\@brx\kern-0.5\wd\@brx\usebox{\@brx}}}
\makeatother

\title{On AdS$_4$ deformations of celestial symmetries}

\author[a]{Roland Bittleston,}
\author[b]{Giuseppe Bogna,}
\author[c]{Simon Heuveline,}
\author[b]{Adam Kmec,}
\author[b]{Lionel Mason}
\author[c]{and David Skinner}

\affiliation[a]{Perimeter Institute for Theoretical Physics,\\ 31 Caroline Street North, Waterloo, Ontario, Canada\vspace{0.1cm}}
\affiliation[b]{The Mathematical Institute, University of Oxford,\\ Woodstock Road, Oxford OX2 6GG, United
Kingdom\vspace{0.1cm}}
\affiliation[c]{Department of Applied Maths \& Theoretical Physics, University of Cambridge,\\ Wilberforce Road, Cambridge CB3 0WA, United Kingdom \vspace{0.1cm}}

\emailAdd{rbittleston@perimeterinsitute.ca}
\emailAdd{giuseppe.bogna@maths.ox.ac.uk}
\emailAdd{sph48@cam.ac.uk}
\emailAdd{adam.kmec@maths.ox.ac.uk}
\emailAdd{lmason@maths.ox.ac.uk}
\emailAdd{d.b.skinner@damtp.cam.ac.uk}

\abstract{
Celestial holography has led to the discovery of new symmetry algebras arising from the study of collinear limits of perturbative gravity amplitudes in flat space.  We explain from the twistor perspective how a non-vanishing cosmological constant $\Lambda$ naturally modifies the celestial chiral algebra. The cosmological constant deforms the Poisson bracket on twistor space, so the corresponding deformed algebra of Hamiltonians under the new bracket is automatically consistent. This algebra is equivalent to that recently found by Taylor and Zhu. We find a number of variations of the deformed algebra.  We give the Noether charges arising from the expression of this algebra as a symmetry of the twistor action for self-dual gravity with cosmological constant.
}

\begin{document}
	\maketitle
    \flushbottom

	
\section{Introduction}

One of the successes of celestial holography \cite{Pasterski:2021raf} has been the identification of new  symmetry algebras of perturbative amplitudes in flat space~\cite{Guevara:2021abz, Strominger:2021lvk}.  It is important to understand how widely such new celestial chiral algebras apply, both in terms of their appearance in different theories on flat space and in terms of whether they exist beyond flat space. In particular, if celestial holography is to be thought of as some limit of conventional AdS/CFT, could these symmetries have some precursor there and, if so, what role might they play in that context?  An answer to this question has been provided by Taylor \& Zhu in~\cite{Taylor:2023ajd}, at least as a first-order deformation in the cosmological constant $\Lambda$. In this work, we show that 
their answer naturally extends   to all orders in $\Lambda$ in a natural  twistorial representation of these symmetries as local holomorphic Hamiltonian diffeomorphisms of twistor space; these have a natural action  on the self-dual sector of Einstein gravity.

\medskip

In flat space, the celestial chiral algebra was first introduced~\cite{Guevara:2021abz} by examining collinear limits and splitting functions of amplitudes in gravity and Yang-Mills  at null infinity $\scri$. A further transform~\cite{Strominger:2021lvk} reveals that the celestial chiral algebra associated to positive helicity gravitons is\footnote{In much of the celestial holography literature, the algebra is called the wedge subalgebra of $Lw_{1+\infty}$. This name is somewhat inaccurate; the true flat space algebra $\LH$ contains the wedge subalgebra of $Lw_{1+\infty}$ as a subalgebra; see {\it e.g.}~\cite{Bittleston:2022jeq,Bittleston:2023bzp} for details.} $\LH$, the loop algebra of Hamiltonian vector fields on $\C^2$. In \cite{Adamo:2021lrv, Mason:2022hly}, it was explained how this transform essentially inverts the Penrose transform~\cite{Penrose:1969ae, Eastwood:1981jy} to realize positive helicity gravitons as deformations of twistor space.  The role of $\LH$ on twistor space was first identified by Penrose \cite{Penrose:1976js, Penrose:1976jq} in his non-linear graviton construction. This gives a correspondence between self-dual vacuum metrics and curved twistor spaces $\mathcal{PT}$.  Locally,  curved twistor spaces are deformations of a region in flat twistor space $\PT$. These deformations are not arbitrary, but are required to preserve a degenerate Poisson structure\footnote{Taking values in $\cO(-2)$, the square root of the canonical bundle, \emph{i.e.}, of homogeneity degree $-2$.} $\{\cdot,\,\cdot\}$, defined by a choice of skew bi-twistor $I^{AB}$ known as the \emph{infinity twistor}. (The reason for the name will become apparent below.) The Poisson algebra of Hamiltonians\footnote{Taking values in $\cO(2)$, the square root of the anti-canonical bundle.} preserving $\{\cdot,\,\cdot\}$ can be readily identified with $\LH$ and the algebra has a two-fold role in this correspondence \cite{Adamo:2021lrv,Mason:2022hly}.  Firstly, it arises as those local holomorphic diffeomorphisms of twistor space that preserve the global geometric structure. Such local symmetries have a second interpretation as defining infinitesimal deformations of the complex structure on $\mathcal{PT}$; the non-linear graviton construction then realizes these as self-dual gravitons on space-time. 

\medskip
  
The non-linear graviton construction was extended to incorporate a cosmological constant by Ward \cite{Ward:1980am} - see also \cite{LeBrun:1982vjh, Salamon:1982}. 
When the cosmological constant is non-vanishing, the non-linear graviton simply relaxes the degeneracy requirement on the infinity twistor and the Poisson structure $\{\cdot,\,\cdot\}_\Lambda$ becomes non-degenerate with an additional term that can be chosen to be proportional to $\Lambda$. 

In this paper, our first aim is to show that the deformation of $\LH$ found by~\cite{Taylor:2023ajd} is indeed the Poisson algebra  $\LHL$ of Hamiltonians for this non-degenerate Poisson bracket on a region $\C^2\times \C^*\subset \PT$. Specifically, the algebra is given by
\begin{equation}
	\{w^p_{m,a},w^{q}_{n,b}\}_\Lambda=(m(q-1)-n(p-1))w^{p+q-2}_{m+n,a+b}-\Lambda(a(q-2)-b(p-2))w^{p+q-1}_{m+n,a+b}\,,
\end{equation}  
as we derive in equation~\eqref{eq:alg-neq0} below. As in the $\Lambda=0$ case, the algebra may be interpreted as both describing infinitesimal diffeomorphisms of twistor space which preserve the Poisson structure, and as the Penrose transform of linearized gravitons. We also point out that different (non-isomorphic) algebras are possible depending on the model of Euclidean AdS$_4$ one considers as this changes the subset $\C^2\times \C^*\subset\PT$. The above algebra is adapted to hyperbolic space being presented as a ball. If one instead uses that upper-half space presentation, a different version of the algebra is obtained. 

The Hamiltonians considered above generate symmetries of the twistor action for self-dual gravity with cosmological constant, first constructed in~\cite{Mason:2007ct}. We will obtain the corresponding Noether charges directly in twistor space, showing that they reduce on-shell to pure boundary terms. This places our work as part of a long tradition of the study of hidden symmetries of the self-duality equations, together with associated conserved quantities, hierarchies and their Hamiltonian origins; see \emph{e.g.}~\cite{Boyer:1985aj,Park:1989fz,Park:1989vq,Mason:1990,Dunajski:2000iq,Dunajski:2003gp} for self-dual gravity and~\cite{Forgacs:1980su,Chau:1981gi,Mason:1991rf} for self-dual Yang Mills.  

\medskip

This work is organized as follows: in section~\ref{sec:TwistorsforAdS} we review the basic construction of the twistor space of AdS$_4$, highlighting the role of the infinity twistor. In Section~\ref{sec:graviton_symmetries}, we explain how one can think of the celestial chiral algebra as the algebra of holomorphic symmetries of the complex structure on twistor space. We show explicitly how the cosmological constant deforms the standard $\LH$ algebra. In section~\ref{sec:symm} we explain how the symmetries arise as symmetries of the twistor action of \cite{Mason:2007ct}  for the self-dual Einstein sector  and identify the associated charges on twistor space.  We conclude in Section \ref{sec:concl} by discussing some open questions and further directions.


\section{Twistors for AdS\texorpdfstring{$_4$}{4}}
\label{sec:TwistorsforAdS}

The twistor space $\PT$ of conformally compactified Minkowski space-time $\M$ is $\CP^3$, often described by homogeneous coordinates $Z^A=(\mu^{\dot\alpha},\lambda_\alpha)$ defined  modulo the $\C^*$ action $Z^A\sim tZ^A$ for $t\in\C^*$. The correspondence between space-time and twistor space is encoded in the incidence relations: points of space-time correspond to holomorphically embedded rational lines $L_x\cong\CP^1\subset\CP^3$.  Explicitly, a standard affine patch of the compactified space-time is given by 
 \begin{equation}
\mu^{\dot\alpha}=\im \,x^{\dot\alpha\alpha}\lambda_\alpha\,,
\label{eq:incidence}
 \end{equation}
with points in real Minkowski space corresponding to choosing $x^{\dot\alpha\alpha}$ Hermitian.

 The construction of twistor space $\PT$ is conformally invariant. Indeed, $\CP^3$ naturally carries an action of $\SL(4,\C)$, the spin group of the complexification of the conformal group $\SO(2,4)$ in four dimensions, acting linearly on the homogeneous coordinates $Z^A$. We can describe the incidence relations more invariantly using embedding space coordinates: we first notice that we can pick a rational line by choosing any pair of distinct twistors $Z_1,Z_2$. Thus we can coordinatize space-time points in terms of skew bitwistors of the form $X^{AB}= Z_1^{[A}Z_2^{B]}$. Conversely, any bitwistor $X^{AB}$ is of this form for some $Z_1,Z_2$ if it satisfies the simplicity condition
\begin{equation}
    X\cdot X\coloneqq \varepsilon_{ABCD}X^{AB}X^{CD}=0\,.
\end{equation}
Skew bitwistors are natural homogeneous coordinates on $\CP^5$, so this construction identifies complexified, conformally compactified Minkowski space as the quadric $\{X\cdot X=0\}\subset\CP^5$. In this sense, the skew bitwistors $X^{AB}$ are embedding coordinates. In terms of these, the incidence relations simply read
\begin{equation}
    \varepsilon_{ABCD}Z^AX^{BC}=0\,,
\end{equation}
which ensures that $Z$ lies on the line defined by the simple $X$.

As is standard, we will let $\mathcal{O}(n)\to\CP^3$ denote the holomorphic line bundle whose sections are functions of homogeneity $n$ in $Z^A$. Provided $\lambda_\alpha\neq0$, the $\lambda_\alpha$ are homogeneous coordinates on $\CP^1$ and we similarly denote the holomorphic line bundle whose sections are functions of homogeneity $n$ in $\lambda_\alpha$ by $\mathcal{O}(n)\to\CP^1$; the distinction between these bundles will be clear from the context.  


\subsection{The infinity twistor and Poisson structure}

A conformal scale is encoded by a choice of skew bitwistor $I^{AB}_\Lambda$ known as the \emph{infinity twistor}.  As our notation emphasises, this bitwistor depends on the cosmological constant $\Lambda$, and is normalized to obey
\begin{equation}
			I^{AB}_{\Lambda}I^\Lambda_{CB}=\Lambda\,\delta^A_C\,,
\label{eq:non-degenerate}
\end{equation}
where $I_{AB}^\Lambda=\frac{1}{2}\varepsilon_{ABCD}I_\Lambda^{CD}$ is the dual of $I^{AB}_\Lambda$. In particular this implies (and in fact is implied by $I^{AB}_\Lambda I^{\Lambda}_{AB} = 4\Lambda$) that only when the cosmological constant is zero does the infinity twistor define a line $L_I\subset\PT$, or a point in space-time. A standard representation is
\begin{subequations}
\begin{equation}
I^{AB}_\Lambda=\left(\begin{array}{cc}
     \varepsilon^{\dot\alpha\dot\beta}&0  \\
     0&\Lambda\,\varepsilon_{\alpha\beta} 
\end{array}\right)
\end{equation}
for the infinity twistor, or equivalently 
\begin{equation}
I^\Lambda_{AB}=\left(\begin{array}{cc}
     \Lambda\,\varepsilon_{\dot\alpha\dot\beta}&0  \\
     0&\varepsilon^{\alpha\beta} 
\end{array}\right)
\end{equation}
\label{eq:infinity_twistors}
\end{subequations} for its dual. Using the infinity twistor, we may define an $\cO(-2)$-valued holomorphic Poisson structure
\begin{equation}
\label{eq:Lambdabracket}
    \{f,g\}_\Lambda :=I_\Lambda^{AB}\frac{\p f}{\p Z^A}\frac{\p g}{\p Z^B}=\varepsilon^{\dot\alpha\dot\beta}\frac{\p f}{\p\mu^{\dot\alpha}}\frac{\p g}{\p\mu^{\dot\beta}}+\Lambda\,\varepsilon_{\alpha\beta}\frac{\p f}{\p\lambda_\alpha}\frac{\p g}{\p\lambda_\beta}\,,
\end{equation}
where the second equalities follow from using the representation
\eqref{eq:infinity_twistors}. Dually, we can use $I^\Lambda_{AB}$ to define the $\cO(2)$-valued 1-form
\begin{equation}
    \tau_\Lambda\coloneqq I^{\Lambda}_{AB}\,Z^A\d Z^B=\langle\lambda\,\d\lambda\rangle+\Lambda[\mu\,\d\mu]\,.
\label{eq:taudef}
\end{equation}
We can state the non-degeneracy condition~\eqref{eq:non-degenerate} more geometrically as follows. Let $\Omega$ be the $\cO(4)$-valued holomorphic volume form
\begin{equation}
\   \Omega\coloneqq\frac{1}{4!}\varepsilon_{ABCD}Z^A\d Z^B\wedge\d Z^C\wedge\d Z^D\,.
\end{equation}
Then~\eqref{eq:non-degenerate} can also be written as  
\begin{equation}
\tau_\Lambda\wedge\d\tau_\Lambda=2\Lambda\,\Omega\,.
\end{equation}
The Poisson structure is then defined as the bivector obtained by contracting $\tau_\Lambda$ with the inverse of $\Omega$.
In this context, $\tau_\Lambda$ is often known as a \emph{contact} form. 

On $\mathbb{C}^2\times \mathbb{C}^*\subset \mathbb{PT}$, we can introduce the inhomogeneous coordinates $(v^{\dot{0}},v^{\dot{1}},z)=(\mu^{\dot{0}}/\lambda_1,\mu^{\dot{1}}/\lambda_1,\lambda_0/\lambda_1)$ in which \eqref{eq:Lambdabracket} takes the form
\be
\label{eq:LambdaBracketInhomogen}
\{f,g\}_\Lambda=\frac{\partial f}{\partial v^{\dot{\alpha}}}   \frac{\partial g}{\partial v_{\dot{\alpha}}} +\Lambda\bigg(v^{\dot{\alpha}} \frac{\partial f}{\partial v^{\dot{\alpha}}}\partial_z g-2f\partial_z g-v^{\dot{\alpha}} \frac{\partial g}{\partial v^{\dot{\alpha}}}\partial_z f+2g\partial_z f\bigg)\,,
\ee
where $f,g$ are sections of $\mathcal{O}(2)$ restricted to $\mathbb{C}^2\times \mathbb{C}^*\subset \mathbb{PT}$, on which the bundle becomes trivial. The two terms $\Lambda(2g\partial_z f-2f\partial_zg)$ resulting from this trivialisation make it manifest that \eqref{eq:LambdaBracketInhomogen} does not obey Leibniz's rule and hence is strictly speaking not a Poisson bracket but rather a so-called \emph{Jacobi bracket}.


\smallskip

Points at the conformal boundary $\scri_\Lambda$ of AdS$_4$ are characterised in embedding space coordinates by  
\begin{equation}
\scri_\Lambda=\{ X| X^{AB}I_{AB}=0\}\,.    \label{infinity}
\end{equation}
In twistor space this is the condition that the restriction of  $\tau_\Lambda$ to the corresponding twistor line vanishes. Using the incidence relations~\eqref{eq:incidence} in~\eqref{eq:taudef}  gives the intersection of $\scri_\Lambda$ with the Euclidean patch as 
\begin{equation}
	\scri_\Lambda=\left\{x^{\alpha\dot\alpha}\in \R^4:  x^2=-2/\Lambda\right\}\, ,
\end{equation}
as expected from the standard form of the AdS metric
\begin{equation}
    g=\frac{\d x^{\alpha\dot\alpha}\d x_{\alpha\dot\alpha}}{\left(1+\frac{1}{2}\Lambda x^2\right)^2}\,.\label{eq:ads_metric0}
\end{equation}
This fixing of the conformal boundary justifies calling $I_\Lambda$ the infinity twistor.

The form~\eqref{infinity} represents Euclidean AdS$_4$ as the interior of a ball of radius $\sqrt{-2/\Lambda}$.  We can instead present the infinity twistors so that Euclidean AdS$_4$ is represented as an upper half space in Poincaré coordinates. Introduce  a constant vector $z^{\alpha\dot\alpha}\coloneqq o^\alpha\bar o^{\dot\alpha}-\iota^\alpha\bar\iota^{\dot\alpha}$ of length $\sqrt{2}$ so that the vertical coordinate of the upper half space representation of $\AdS_4$ can be defined to be $z\coloneqq x^{\alpha\dot\alpha}z_{\alpha\dot\alpha}/\sqrt{2}$ on which the metric becomes
\begin{equation}
    g=\frac{\d x^{\alpha\dot\alpha}\d x_{\alpha\dot\alpha}}{\Lambda z^2}\, .
\end{equation}
 Eliminating dotted indices via  $\mu^{\alpha}\coloneqq z^\alpha{}_{\dot\alpha}\mu^{\dot\alpha}$, the infinity twistors that reproduce the metric in Poincaré coordinates are
\begin{equation} I^\Lambda_{AB}=I^{AB}_{\Lambda}=\sqrt{\Lambda}\left(\begin{array}{cc}
         0&\varepsilon^{\alpha\beta}  \\
         -\varepsilon_{\alpha\beta}&0 
    \end{array}\right)\,\label{eq:infinity_twistor_poincare}\, .
\end{equation}
This gives 
the Poisson bracket and contact form
\begin{equation}
\begin{aligned}
   \{\cdot,\,\cdot\}^{P}_\Lambda&=\sqrt{\Lambda}\,\frac{\p }{\p\lambda_\alpha}\wedge\frac{\p }{\p \mu^\alpha}\,,\\
   \tau_\Lambda&=\sqrt{\Lambda}\,\langle\mu\,\d\lambda\rangle-\sqrt{\Lambda}\,\langle\d\mu\,\lambda\rangle\,.
\end{aligned}
\end{equation}
For the most part we will use to the ball coordinates \eqref{eq:ads_metric0}, as they make the limit $\Lambda\to0$ to flat space straightforward. 


\subsection{The non-linear graviton with cosmological constant}

Twistor theory extends beyond conformally flat space-times: Penrose's non-linear graviton construction 
\cite{Penrose:1976jq,Penrose:1976js} establishes a correspondence between self-dual (SD) vacuum metrics and certain deformed twistor spaces.  Ward \cite{Ward:1980am} extended the non-linear graviton construction to include non-zero cosmological constant as follows\footnote{See also \cite{LeBrun:1982vjh, Salamon:1982,  Hitchin:1982vry} for extensions and variations of this construction.}

\begin{thm}[Ward '80]\label{th:non-linear_graviton_with_Lambda}
    There is a 1-to-1 correspondence between complex self-dual Einstein manifolds $(M,g)$ and deformations $\mathcal{PT}$ of a neighbourhood of a line in $\PT$ preserving $\{\cdot,\,\cdot\}_\Lambda$ as a Poisson structure with values in $\cO(-2)$, the square root of the canonical bundle.

    A real slice of $M$ of signature $(2,2)$ or $(4,0)$ corresponds to an antiholomorphic involution $\sigma:\mathcal{PT}\rightarrow \mathcal{PT}$ that, for signature $(2,2)$, fixes a real slice $\mathcal{PT}_\R$. For Euclidean signature the involution $\sigma$ has no fixed points. 
\end{thm}
Preservation of the Poisson structure can be formulated dually in terms of $\tau_\Lambda$, which can be defined as the 1-form with values in the square root of the  anti-canonical line bundle dual to the Poisson structure (\emph{i.e.}, obtained by contracting  the Poisson structure into $\Omega$ thought of as a 3-form with values in the anti-canonical bundle).

Briefly, the theorem is proved in the forward direction by constructing the curved twistor space as the space of totally null ASD 2-surfaces in some small complexification; their existence follows from the vanishing of the ASD Weyl curvature.  In Euclidean signature these simply hit the real slice in a unique point with tangent plane defined by an ASD spinor up to scale, so that the twistor space can be identified with the total space of the bundle of projective ASD spinors~\cite{Atiyah:1978wi,Hitchin:1982vry,Salamon:1982}.  In the reverse direction, space-time is realized as the moduli space of degree-1 holomorphic curves in $\mathcal{PT}$; those at $\scri$ are those on which the contact form $\tau_\Lambda$ vanishes.  The real slice $M_\R\subset M$ is given by  those degree-1 holomorphic curves that are sent to themselves by the anti-holomorphic involution. 
 

\section{Symmetries and gravitons}
\label{sec:graviton_symmetries}

The infinitesimal symmetries of AdS$_4$ are naturally defined by holomorphic functions of homogeneity degree 2 on twistor space, given as $h=\frac{1}{2}h_{AB}Z^AZ^B$ for a constant, symmetric $h_{AB}$. The associated Hamiltonian vector fields $X_h= \{h\,,\cdot \}_{\Lambda} =  Z^Ah_{AB}I^{BC}_\Lambda\p_C$  generate the corresponding motions of twistor space. By construction, flows along such Hamiltonian vector fields preserves the Poisson bracket and its dual contact 1-form. In space-time, this motion induces the standard isometries of AdS$_4$, arising as the spin group Sp$(2)$ of the more usual $\SO(3,2)$. These isometries provide the starting-point for the full celestial chiral algebra of AdS$_4$.

More generally, local holomorphic diffeomorphisms that are symmetries of the Poisson structure allow singularities in the Hamiltonians, with  the algebra of such holomorphic symmetries being simply the Poisson structure between two generators. In flat space-time, it was shown in~\cite{Adamo:2021lrv} that this algebra can be identified with the celestial chiral algebra by using the regularity near $\mu^{\dot\alpha}=0$ to decompose such $h$ into modes of the form
\be \label{eq:basis}
    w^{p}_{m,a} = \frac{(\mu^{\dot 0})^{p+m-1}(\mu^{\dot 1})^{p-m-1}}{2\lambda_0^{p-a-2}\lambda_1^{p+a-2}}\,.
\ee
Here the parameters $p,m$ have been chosen to agree with their counterparts in \cite{Strominger:2021lvk} with ranges fixed by the requirement that $p\pm m-1\in \Z_{\geq 0}$. In this range,  negative powers of $\mu^{\dot\alpha}$ do not arise. The parameter $a\in\Z+p$ has been shifted relative to its analog in \cite{Adamo:2021lrv} so as to match the conventions of \cite{Taylor:2023ajd}. It is simple to check that, using the Poisson bracket for $\Lambda=0$, the Poisson algebra of such generators yields the celestial chiral algebra
\be
    \{w^p_{m,a},w^q_{n,b}\}_{\Lambda=0}=(m(q-1)-n(p-1))w^{p+q-2}_{m+n,a+b}\,, \label{eq:alg-0}
\ee
of flat space-time. This algebra was originally derived via a bottom-up calculation of soft symmetry algebras at null infinity \cite{Guevara:2021abz}, followed by a light-ray transform \cite{Strominger:2021lvk}. 

We have denoted this algebra~\eqref{eq:alg-0} by  $\LH$ because it is
the algebra of Hamiltonian functions on the $\C^2$ coordinatized by $\mu^{\dot\alpha}$; for $\Lambda=0$ the Poisson structure defined by the infinity twistor acts only in the $\mu^{\dot\alpha}\in \C^2$ variables and is global in this $\C^2$.
One gets the loop algebra because $\lambda_\alpha$ appear only as parameters that are only required to be holomorphic for $\lambda_0/\lambda_1 \in \C^*\supset S^1$  leading to the loop $L$ in the notation $\LH$.  

\smallskip

The algebra could equivalently be denoted $\LHL|_{\Lambda=0}$ as the generators are holomorphic on  $(\mu^{\dot\alpha}/\lambda_1, \lambda_0/\lambda_1)\in \C^2\times \C^*$
and the $\Lambda=0$ Poisson structure does not see the $\lambda_\alpha$ variables.
The above considerations extend straightforwardly to the case of non-vanishing $\Lambda$. Explicitly, using the $\Lambda$-deformed Poisson bracket $\{\cdot,\,\cdot\}_\Lambda$, the flat space celestial chiral algebra is deformed to\footnote{Note that this Lie algebra deformation does \emph{not} arise as the loop algebra of a deformation of $\mathfrak{ham}(\C^2)$ itself. Indeed, the unique Lie algebra deformation of the latter is the Weyl algebra corresponding to a non-commutative self-dual gravity \cite{Bu:2022iak}.}
\be
	\{w^p_{m,a},w^q_{n,b}\}_\Lambda = (m(q-1)-n(p-1))w^{p+q-2}_{m+n,a+b}-\Lambda(a(q-2)-b(p-2))w^{p+q-1}_{m+n,a+b}\,, \label{eq:alg-neq0}
\ee
by the presence of a cosmological constant.   As above, for $2-p\leq a\leq p-2$ the $w^p_{m,a}$ are the quadratic Hamiltonians generating the standard AdS$_4$ isometries. 

The algebra~\eqref{eq:alg-neq0} agrees with that found by Taylor \& Zhu~\cite{Taylor:2023ajd}, who considered the holomorphic collinear limit of the Mellin transform of the leading order in $\Lambda$ correction to the 4-graviton tree level amplitude (sum of all tree-level Witten diagrams) in AdS$_4$. This Mellin amplitude was computed in~\cite{Alday:2022uxp, Alday:2022xwz, Alday:2023jdk, Alday:2023mvu}. Interestingly, \cite{Taylor:2023ajd} found that the true 4-pt amplitude does \emph{not} quite exhibit this algebra; the form of the $O(\Lambda)$ correction 
to be modified in order to ensure the Jacobi identity holds. For us, the fact that the algebra arises from a Poisson bracket immediately ensures that the Jacobi identity is satisfied. The twistor construction suggests that the algebra~\eqref{eq:alg-neq0} is a true celestial symmetry of \emph{self-dual} gravity on AdS$_4$, at least at the classical level.


\subsection{Elements of \texorpdfstring{$\LHL$}{ham(C2xC*)} as gravitons}

More general singular elements of the celestial chiral algebra correspond to allowing positive helicity gravitons as fluctuations on the background. In twistor space, the construction of Theorem \ref{th:non-linear_graviton_with_Lambda} identifies self-dual gravitons with infinitesimal deformations that are Hamiltonian. In the original \v Cech presentation~\cite{Penrose:1976jq,Penrose:1976js}, these are determined by a cohomology element $[h]\in H^{1}(\PT,\cO(2))$. Specifically, for a \v Cech cohomology description, one covers $\PT$ by two topologically trivial open sets $\PT=U_0\cup U_1$, where $U_0=\{ Z^A\in\PT:\lambda_0\neq 0\} $ and $U_1=\{Z^A\in\PT:\lambda_1\neq 0\}$. The class $[h]$ is then simply represented as a homogeneity $+2$ holomorphic function on the overlap $U_0\cap U_1$. Thus positive-helicity gravitons are equivalent to holomorphic symmetries of the contact structure and Poisson bracket on $U_0\cap U_1$ modulo gauge. 

The gauge modes with $a\leq p-2$ can be extended holomorphically over all of $U_1$, while those with $a\geq2-p$ extend holomorphically over all of $U_0$. In \v Cech cohomology, these modes are pure gauge, generating coordinate transformations (rather than genuine deformations) of $U_1$ and $U_0$ respectively.
Thus, as a symmetry algebra, it is acting on triples consisting of the complex structure of $\mathcal{PT}$ together with  a coordinate charts on $U_0$ and $U_1$.

If $2-p\leq a\leq p-2$, $w^p_{m,a}$ are the quadratic Hamiltonians whose associated vector fields generate the subgroup of the Poincar\'e group consisting of the translations and self-dual rotations.  More generally, for $p=3/2$ these generate certain holomorphic but singular supertranslations, while for $p=2$ they generate self-dual superrotations. (See \cite{Adamo:2021lrv} for further explanation.)


\subsection{Celestial chiral algebras as vertex algebras} \label{subsec:VOA}

Lie algebras of local symmetries on twistor space are closely associated with vertex algebras supported on twistor lines \cite{Adamo:2021lrv,Costello:2022wso}. We can understand the vertex algebra corresponding to \eqref{eq:alg-neq0} in the following way. Suppose we couple the twistor uplift of self-dual gravity on AdS$_4$ to a 2d holomorphic theory living on a twistor line. In general such a coupling will take the  form
\be \label{eq:defect} \sum_{p\pm m\in\mathbb{N}}\frac{2}{(p+m-1)!\,(p-m-1)!}\int_{\CP^1_x}\la\lambda\,\d\lambda\ra\wedge\,w^p_m(\lambda)\,\p_{\mu^{\dot0}}^{p+m-1}\p_{\mu^{\dot 1}}^{p-m-1}H\,\,, 
\ee
for 2d operators $w^p_m(\lambda)$ depending meromorphically on $\lambda$ and labelled by $p,m$ with the same ranges as above. Here $H\in\Omega^{0,1}(\PT,\cO(2))$ is a Dolbeault representative corresponding to the \v Cech cocycle $[h]$. In order for the integrand to have vanishing homogeneity the operators $w^p_m(\lambda)$ must take values in $\cO(2p-6)$, \emph{i.e.}, they must have conformal spin $3-p$.

Our notation for the operators $w^p_m(\lambda)$ can be justified by choosing $H$ corresponding to the basis \eqref{eq:basis}. Explicitly, we fix $H = w^p_{m,a}\,\dbar B$ for $B$ a bump function on $\CP^1_x$ taking the value 1 in a neighbourhood of $\lambda_0=0$, 0 in a neighbourhood of $\lambda_1=1$ and non-constant on an annulus disconnecting $\lambda_0,\lambda_1$. Introducing inhomogeneous coordinates on $\CP^1_x$ such that $\lambda\sim(z,1)$, in the limit of an arbitrarily narrow annulus we may take $B = \Theta(|z|^2<1)$ for $\Theta$ the Heaviside step function. Substituting into \eqref{eq:defect} gives
\be \oint\frac{\la\lambda\,\d\lambda\ra}{\lambda_0^{p-a-2}\lambda_1^{p+a-2}}\,w^p_m(\lambda) = \oint\d z\,z^{a+2-p}w^p_m(z)\,. \ee
In this way the Hamiltonians $w^p_{m,a}$ are naturally identified with the modes of the operators $w^p_m(z)$.

Invariance of the coupling \eqref{eq:defect} under the local holomorphic diffeomorphism symmetry on twistor space necessitates the following operator products
\be
\begin{aligned} \label{eq:VOA-neq0}
&w^p_m(z_1)w^q_n(z_2)
\sim \frac{m(q-1)-n(p-1)}{z_{12}}w^{p+q-2}_{m+n}(z_2) \\
&- \frac{\Lambda}{z_{12}^2}\big((p+q-4)w^{p+q-1}_{m+n}(z_2) + z_{12}(p-2)\p_zw^{p+q-1}_{m+n}(z_2)\big)\,.
\end{aligned}
\ee
This is the (tree level) celestial chiral algebra of SD gravity on AdS$_4$ represented as a vertex algebra. We remark that the field $T(z) = w^1_0(z)/\Lambda$ plays the role of a stress tensor, with OPEs
\be T(z_1)w^p_m(z_2) \sim \frac{1}{z_{12}^2}\big((3-p)w^p_m(z_2) + z_{12}\p_zw^p_m(z_2)\big)\,. \ee
The corresponding central charge vanishes. Furthermore, $T(z)$ is the field of conformal spin 2 in a vertex subalgebra generated by $w^p_0(z)/\Lambda$ for $p\in\Z_{\geq1}$. This resembles the $w_\infty$ vertex algebra which has the same defining operator products, but for labels taking values in the range $p\in\Z_{\leq1}$. $w_\infty$ is generated by fields of all integer conformal spins $s\geq2$ rather than $s\leq2$.

The fields $w^p_m(z)$ can be conveniently organised into hard generating functions depending on an auxiliary right-handed spinor $\tilde\lambda_{\dot\alpha}$, defined by
\be w(\tilde\lambda,z) = \sum_{p\pm m\in\Z_{\geq1}}\frac{2\tilde\lambda_{\dot0}^{p+m-1}\tilde\lambda_{\dot1}^{p-m-1}}{(p+m-1)!(p-m-1)!}w^p_m(z)\,. \ee
In terms of these hard generators the vertex algebra \eqref{eq:VOA-neq0} reads
\be
\begin{aligned}
&w(\tilde\lambda_1,z_1)w(\tilde\lambda_2,z_2) \\
&\sim \frac{[12]}{z_{12}}w(\tilde\lambda_1+\tilde\lambda_2,z_2) - \frac{\Lambda}{z_{12}^2}\Big(\big(\tilde\lambda_{\dot\alpha}\p_{\tilde\lambda_{\dot\alpha}} - 4\big)w(\tilde\lambda,z_2) + z_{12}\big(\tilde\lambda_{1\dot\alpha}\p_{\tilde\lambda_{\dot\alpha}} - 2\big)\p_zw(\tilde\lambda,z_2)\Big)\Big|_{\tilde\lambda = \tilde\lambda_1+\tilde\lambda_2}\,.
\end{aligned}
\ee
We recognise the coefficient $[12]/z_{12}$ as the tree graviton splitting function on flat space \cite{Bern:1998xc}. But the $\Lambda$ dependent coefficients are not simply functions of the spinor-helicity variables, instead taking the form of differential operators. This results from the loss of supertranslation invariance on AdS$_4$.


\subsection{Variations and extensions}

Had we used the representation of the infinity twistor \eqref{eq:infinity_twistor_poincare} appropriate to the Poincar\'e patch, we'd have obtained the algebra
\be
\begin{aligned} \label{eq:Poincare-alg}
    &\{\hat w^p_{m,a},\hat w^q_{n,b}\}_\Lambda^P = \big((p+m-1)(q-b-2) - (q+n-1)(p-a-2)\big)\hat w^{p+q-3/2}_{m+n-1/2,a+b-1/2} \\
    &+ \big((p-m-1)(q+b-2) - (q-n-1)(p+a-2)\big)\hat w^{p+q-3/2}_{m+n+1/2,a+b+1/2}\,,
\end{aligned}
\ee
where the generators of the algebra have been redefined to 
\be
  \hat w^{p}_{m,a}  \coloneqq \frac{2}{\sqrt{\Lambda}}w_{m,a}^{p},
\ee
Although these again provide an extension of the $\AdS_4$ symmetries, they are not suitable for expansion around $\Lambda=0$.  They are however well adapted to soft limits for momentum eigenstates based on translations of the Poincar\' e patch. We emphasise that the Lie algebra \eqref{eq:Poincare-alg} is \emph{not} isomorphic to \eqref{eq:alg-neq0}. The difference is  essentially the choice of the location of the line $\lambda_\alpha=0$: up to AdS$_4$ isometries there are two such choices, the first where the line is in the complement of the unit ball, and the second here where the line corresponds to a point of $\scri$. These two choices provide the two algebras~\eqref{eq:alg-neq0} \&~\eqref{eq:Poincare-alg}, respectively. On the other hand, the choices of the sets $U_0$ and $U_1$ used to define our basis $\{w^p_{m,n}\}$ are more associated to the choice of cohomology representation. Indeed, these can be made canonically in split signature.

Following the steps outlined in section~\ref{subsec:VOA}, one may recover the vertex algebra associated to~\eqref{eq:Poincare-alg}, which reads
\be
\begin{aligned}
    &w^p_m(z_1)w^q_n(z_2) \sim \frac{(p+q+m+n-2)}{z_{12}^2}\hat w^{p+q-3/2}_{m+n-1/2}(z_2) + \frac{(p+m-1)}{z_{12}}\p_z\hat w^{p+q-3/2}_{m+n-1/2}(z_2) \\
    &- \frac{(p+q-m-n-2)}{z_{12}^2}z_2\hat w^{p+q-3/2}_{m+n+1/2}(z_2) - \frac{(p-m-1)}{z_{12}}\p_z(z_2\hat w^{p+q-3/2}_{m+n+1/2}(z_2)) \\
    &+ \frac{2\big((p-2)(n-1) - (q-2)(m-1)\big)}{z_{12}}\hat w^{p+q-3/2}_{m+n+1/2}(z_2)\,.
\end{aligned}
\ee
It's straightforward to recast the above in terms of hard generators, which we leave as an exercise for the interested reader. 

\smallskip

Both the Lie algebra adapted to the ball \eqref{eq:alg-neq0} and the Poincar\'{e} patch \eqref{eq:Poincare-alg} can easily be extended to incorporate free fields propagating on the gravitational background. By the linear Penrose transform, solutions to free field equations of spin $s$ on AdS$_4$ are in bijection with cohomology classes $[\varphi]\in H^1(\PT,\cO(2s-2))$. Fluctuations of such fields can be represented in the \v Cech language by holomorphic functions on $U_0\cap U_1$, with homogeneity $2s-2$. Regularity near $\mu^{\dot\alpha}=0$ allows us to decompose $\varphi$ into modes
\be x^p_{m,a} = \frac{(\mu^{\dot0})^{p+m-1}(\mu^{\dot1})^{p-m-1}}{2\lambda_0^{p-a-s}\lambda_1^{p+a-s}} \ee
where as above $p\pm m-1\in\Z_{\geq0}$. The Hamiltonians $w^p_{m,a}$ naturally act on these modes via the Poisson bracket, furnishing us with modules for the Lie algebras $\eqref{eq:alg-neq0}$ and \eqref{eq:Poincare-alg}. For example, in the case of the ball model this action reads
\be \label{eq:spin-s-module}
	\{w^p_{m,a},x^q_{n,b}\}_\Lambda = (m(q-1)-n(p-1))x^{p+q-2}_{m+n,a+b}-\Lambda(a(q-s)-b(p-2))x^{p+q-1}_{m+n,a+b}\,.
\ee
Extending by these modules gives symmetry algebras for the coupled systems. As we can see in equation \eqref{eq:spin-s-module}, for $\Lambda\neq0$ the structure constants of such extended algebras depend explicitly on $s$. This reflects the fact that for general fluctuations $h$ the resulting curved twistor space does not holomorphically fibre over $\CP^1$. The complex structure of the line bundles $\cO(2s-2)$ is deformed to $\mathcal{K}^{(1-s)/2}$ for $\mathcal{K}$ the canonical bundle of the curved twistor space. The $s$-dependent term in \eqref{eq:spin-s-module} is generated by this shift.

To incorporate self-dual Yang-Mills is no more difficult. Conformal invariance ensures that the $S$-algebra is undeformed on AdS$_4$, and the natural action of \eqref{eq:alg-neq0} and \eqref{eq:Poincare-alg} outlined above distributes over it's commutators. Therefore we can extend in the same way.

It is also easy to extend these considerations to include supersymmetry. We adjoin  $\mathcal{N}$ fermionic  coordinates $\eta^I$ to give the homogeneous coordinates $\cZ^{\cI}\coloneqq (Z^A,\eta^I)$ on $\CP^{3|\cN}$, acted on by the superconformal group $\SL(4|\mathcal{N})$.  As shown in~\cite{Wolf:2007tx, Mason:2007ct}, self-dual $\SO(\cN)$ gauged supergravity on AdS$_4$ may be described by the non-linear graviton construction, where the infinity twistor is extended to a non-degenerate graded skew supertwistor $I^{\cI\cJ}$ defining graded Poisson structure and contact form. This preserves a $\mathrm{OSP}(2|\cN)$ subgroup of the superconformal group. Ungauged supergravities can be obtained by considering infinity twistors that do not have maximal rank in the fermionic directions. In particular, the fully ungauged supergravity arises when we continue to use the non-supersymmetric Poisson structure. By augmenting the $w^p_{m,a}$ to a basis of functions of the homogeneous supertwistor coordinates $\cZ^{\cI}$ of homogeneity 2, we obtain supersymmetric extensions of the above algebras. These provide extensions of the relevant super-Lie algebras of symmetries for supersymmetric extensions of $\AdS_4$ depending the choice of infinity twistor. They also includes twistor functions for all the supergravity modes.


\section{Symmetries of the twistor action}\label{sec:symm}

These symmetries can be understood via Noether arguments applied to the twistor action. We give a brief sketch here. Self-dual gravity with $\Lambda\neq0$ may be described on twistor space by the Poisson-BF theory  
\begin{equation}
    S_{\text{PBF}}=\int_\PT\Omega\wedge G \wedge \left(\dbar H+\frac{1}{2}\{H,H\}_\Lambda\right)\,,\label{eq:poisson-bf}
\end{equation}
first introduced in~\cite{Mason:2007ct}. Here, $\Omega$ is the $\mathcal{O}(4)$-valued top holomorphic form on twistor space, while $H\in\Omega^{0,1}(\PT,\mathcal{O}(2))$ is a Hamiltonian governing the complex structure deformation on twistor space via the the deformed Dolbeault operator
\begin{equation}
    \bar\nabla:=\dbar+\{H,\,\cdot\,\}_\Lambda.
\end{equation} Finally, $G\in\Omega^{0,1}(\PT,\mathcal{O}(-6))$ is a Lagrange multiplier that, on-shell, may be interpreted as the Penrose transform of a linearized ASD Weyl spinor propagating on the SD background.  

\medskip 

The celestial chiral algebra above is related to  gauge transformations of this Lagrangian.  These are a combination of Poisson diffeomorphisms of twistor space, generated by the Hamiltonian vector field associated to a smooth function $\chi$ of homogeneity degree 2, and a further transformation generated by the Hamiltonian vector field of a smooth function $\tilde\chi $ of weight $-6$. These transformations act on the fields as
\begin{equation}
    \delta H= \bar\nabla \chi\, , \qquad \delta G= \bar\nabla \tilde\chi +\{\chi ,G\}_\Lambda\,.
\label{eq:gauge-transforms}
\end{equation}

To make contact with the celestial chiral algebra, we compute the Noether charges associated to these field transformations. The (pre-)symplectic form on the space of classical solutions of the twistor action~\eqref{eq:poisson-bf} can easily be seen to be
\begin{equation}
    \omega(\delta H,\delta G )=\int_{\Sigma} \delta H\wedge \delta G \wedge \Omega \,,
\end{equation}
where $\Sigma$ is real co-dimension 1 slice of $\PT$. The Noether charges corresponding to~\eqref{eq:gauge-transforms} are then given by the integrals 
\begin{equation}
\begin{aligned}
\mathcal{H}_{\chi} &= \int_\Sigma \bar\nabla\chi \wedge G \wedge \Omega\\
\mathcal{H}_{\tilde\chi} &= \int_\Sigma \bar\nabla \tilde \chi\wedge H\wedge\Omega\,,
\end{aligned}
\end{equation}
On shell, after integrating by parts these reduce to boundary terms
\begin{equation}
\begin{aligned}
\mathcal{H}_{\chi}&= \int_{\p\Sigma} \chi\,  G \wedge\Omega\\
\mathcal{H}_{\tilde\chi} &=  \int_{\p\Sigma} \tilde \chi\, H\wedge\Omega\, .
\end{aligned}
\label{eq:boundaryNoether}
\end{equation}
as expected for Noether charges of gauge transformations. The real four-manifold  $\p\Sigma$ can for example be taken to be the 4-surface swept out by Riemann spheres $L_x$ as $x$ varies over some choice of 2-surface in space-time. In this case, the Penrose transform may be used to express the integrals~\eqref{eq:boundaryNoether} in terms of fields on space-time.

The Noether charge integrals~\eqref{eq:boundaryNoether} are not themselves gauge invariant unless $\chi, \tilde \chi$ are holomorphic.  Such global holomorphic $\chi$ of homogeneity degree two are, as described above, the Hamiltonians that generate the global isometries of AdS$_4$. Because $\tilde \chi $ has weight $-6$, there are no such global $\tilde \chi$. If we wish to allow singularities so as to extend the $\chi$ to be the generators \eqref{eq:basis} of our extended algebra, we must  extend our on-shell  phase space by imposing boundary conditions so that $H=G=0$ near $\lambda_0=0$ and $\lambda_1=0$.  This is equivalent to choosing holomorphic Darboux coordinates on small neighbourhoods of $\lambda_0=0$ and $\lambda_1=0$ as discussed earlier. The Noether charges~\eqref{eq:boundaryNoether} make sense on this extended phase space and generate the celestial chiral algebra for $\Lambda\neq0$ described earlier.  


\section{Conclusion and discussion}\label{sec:concl}

We have seen that the first order deformation to the flat space celestial chiral algebra found in \cite{Taylor:2023ajd} naturally arises from local Poisson diffeomorphisms of twistor space, once the flat-space Poisson structure is replaced by the Poisson structure defined by a non-degenerate infinity twistor. In \cite{Taylor:2023ajd}, the algebra was constructed via the AdS amplitudes first found in \cite{Alday:2022uxp, Alday:2022xwz, Alday:2023jdk, Alday:2023mvu}, although in a rather \emph{ad hoc} manner: the algebra that arises na\"ively from the AdS amplitudes fails to satisfy the Jacobi identity, so a suitable modification of the $\Lambda$-deformed graviton OPE is needed to restore associativity. In contrast, here we have constructed the algebra from first principles directly from the Poisson bracket on twistor space, so the Jacobi identity is automatically satisfied. We thus see that this algebra is the celestial chiral algebra of self-dual gravity of AdS$_4$, at least at the classical level.  We have seen further that the algebra can be understood as symmetries of an extension of the twistor action for the SD Einstein equations leading to Noether charges on an extended phase space. 

\medskip

Perhaps the most interesting question is the extent to which these symmetries can yield useful insights beyond the self-dual sector. However, many more modest  directions deserve to be explored in further work:

\paragraph{Twistor sigma models.} The chiral sigma models of \cite{Skinner:2013xp,Mason:2022hly} naturally extend to ones with a non-degenerate infinity twistor. They represent the $\Lambda$-deformed chiral algebra directly in the OPE of their vertex operators.  Correlation functions of these vertex operators naturally then give rise to the split signature correlators for Einstein gravity.

\paragraph{Perturbiner calculations.} In flat space, the celestial chiral algebra can be seen in the splitting function of gravity amplitudes. This is essentially a pertubiner: two on-shell positive helicity states joined to a propagator at a trivalent vertex, taken in the limit that the external momentum become (holomorphically) collinear. It would be interesting to perform these calculations in AdS$_4$, with bulk-to-boundary propagators as external states. This would bring together the perspective of~\cite{Taylor:2023ajd} with the current work, albeit in the context of self-dual gravity.

\paragraph{Space-time realizations and AdS/CFT.}  Here we have focused on formulations in twistor space.  There are by now a number of space-time formulations of many of these ideas at $\Lambda=0$ such as \cite{Freidel:2021ytz,Donnay:2024qwq} and it would be interesting to extend these to $\Lambda\neq 0$, perhaps following on from the frameworks developed by \cite{Hoegner:2012sq,Lipstein:2023pih, Neiman:2023bkq}.  In a different direction, it will be interesting to identify the self-dual sectors of conventional AdS$_4$/CFT$_3$ correspondences such as, for example, those described in \cite{Aharony:2008ug}, so as to see how these structures arise there.    There are also a number of other $\Lambda$-BMS proposals to be compared to from a space-time perspective such as \cite{Compere:2019bua,Compere:2020lrt,Fiorucci:2020xto}.

\paragraph{Further deformations with $\Lambda \neq 0$.}
There is by now a fairly large number of works on classical deformations of celestial chiral algebras (or the absence thereof) in asymptotically (locally) flat space-times \cite{Bu:2022iak, Bittleston:2023bzp, Adamo:2023zeh, Mago:2021wje}. All of these are loop algebras of deformations of $\mathfrak{ham}(\mathbb{C}^2)$ or $\mathfrak{ham}(\mathbb{C}^2/\Gamma)$ for some finite subgroup $\Gamma \subset \SU(2)$ which are very restricted. The Lie algebra \eqref{eq:alg-neq0} in turn is not the loop algebra of a deformation of $\mathfrak{ham}(\mathbb{C}^2)$. It is natural to ask whether there are further deformations of this form and what their geometric bulk-interpretations are. Gravitational instantons in the presence of a cosmological constant are currently being investigated \cite{Bogna:2024soon}. 

\paragraph{Quantum corrections.} For $\Lambda=0$ Poisson-BF theory on twistor space is anomalous \cite{Bittleston:2022nfr}, signalling a loss of integrability in SD gravity at the quantum level. Certainly we should expect a similar anomaly in Poisson-BF theory for $\Lambda\neq0$. It would be interesting to compute this, particularly to see whether there exist alternative methods of anomaly cancellation. Successful cancellation of the anomaly would presumably lead to consistent quantum counterparts of the tree level celestial chiral algebra \eqref{eq:VOA-neq0} along the lines of \cite{Costello:2022upu,Bittleston:2022jeq}.


\acknowledgments  We thank Tim Adamo, Sean Seet and Atul Sharma, to whom these ideas will be well-known, and Romain Ruzziconi, Akshay Yelleshpur Srikant for encouragement and useful discussions on these and related topics. RB and SH would like to thank Kevin Costello and Natalie Paquette for helpful discussions. This work was supported by the Simons Collaboration on Celestial Holography. RB's research at Perimeter Institute is supported in part by the Government of Canada through the Department of Innovation, Science and Economic Development and by the Province of Ontario through the Ministry of Colleges and Universities. GB is supported by a joint Clarendon Fund and Merton College Mathematics Scholarship. SH is partly supported by St. John’s College, Cambridge. AK is supported by the STFC.  LJM would like to thank the STFC for financial support from grant number ST/T000864/1.  The work of SH \& DS has been supported in part by STFC consolidated grant ST/X000664/1.


	\bibliography{eads4_w1infinity}

\providecommand{\href}[2]{#2}\begingroup\raggedright\begin{thebibliography}{10}

\bibitem{Pasterski:2021raf}
S.~Pasterski, M.~Pate, and A.-M. Raclariu, {\it {Celestial Holography}},  in {\em {Snowmass 2021}}, 11, 2021.
\newblock \href{http://arxiv.org/abs/2111.11392}{{\tt arXiv:2111.11392}}.

\bibitem{Guevara:2021abz}
A.~Guevara, E.~Himwich, M.~Pate, and A.~Strominger, {\it {Holographic Symmetry Algebras for Gauge Theory and Gravity}},  \href{http://arxiv.org/abs/2103.03961}{{\tt arXiv:2103.03961}}.

\bibitem{Strominger:2021lvk}
A.~Strominger, {\it {w(1+infinity) and the Celestial Sphere}},  \href{http://arxiv.org/abs/2105.14346}{{\tt arXiv:2105.14346}}.

\bibitem{Taylor:2023ajd}
T.~R. Taylor and B.~Zhu, {\it {w(1+infinity) Algebra with a Cosmological Constant and the Celestial Sphere}},  \href{http://arxiv.org/abs/2312.00876}{{\tt arXiv:2312.00876}}.

\bibitem{Bittleston:2022jeq}
R.~Bittleston, {\it {On the associativity of 1-loop corrections to the celestial operator product in gravity}},  {\em JHEP} {\bf 01} (2023) 018, [\href{http://arxiv.org/abs/2211.06417}{{\tt arXiv:2211.06417}}].

\bibitem{Bittleston:2023bzp}
R.~Bittleston, S.~Heuveline, and D.~Skinner, {\it {The Celestial Chiral Algebra of Self-Dual Gravity on Eguchi-Hanson Space}},  \href{http://arxiv.org/abs/2305.09451}{{\tt arXiv:2305.09451}}.

\bibitem{Adamo:2021lrv}
T.~Adamo, L.~Mason, and A.~Sharma, {\it {Celestial $w_{1+\infty}$ Symmetries from Twistor Space}},  {\em SIGMA} {\bf 18} (2022) 016, [\href{http://arxiv.org/abs/2110.06066}{{\tt arXiv:2110.06066}}].

\bibitem{Mason:2022hly}
L.~Mason, {\it {Gravity from holomorphic discs and celestial $Lw_{1+\infty }$ symmetries}},  {\em Lett. Math. Phys.} {\bf 113} (2023), no.~6 111, [\href{http://arxiv.org/abs/2212.10895}{{\tt arXiv:2212.10895}}].

\bibitem{Penrose:1969ae}
R.~Penrose, {\it {Solutions of the zero-rest-mass equations}},  {\em J. Math. Phys.} {\bf 10} (1969) 38--39.

\bibitem{Eastwood:1981jy}
M.~G. Eastwood, R.~Penrose, and R.~O. Wells, {\it {Cohomology and Massless Fields}},  {\em Commun. Math. Phys.} {\bf 78} (1981) 305--351.

\bibitem{Penrose:1976js}
R.~Penrose, {\it {Nonlinear gravitons and curved twistor theory}},  {\em Gen. Rel. Grav.} {\bf 7} (1976) 31--52.

\bibitem{Penrose:1976jq}
R.~Penrose, {\it {The Nonlinear Graviton}},  {\em Gen. Rel. Grav.} {\bf 7} (1976) 171--176.

\bibitem{Ward:1980am}
R.~S. Ward, {\it {Self-dual space-times with cosmological constant}},  {\em Commun. Math. Phys.} {\bf 78} (1980) 1--17.

\bibitem{LeBrun:1982vjh}
C.~R. LeBrun, {\it {\ensuremath{\mathscr{H}}-Space with a Cosmological Constant}},  {\em Proc. Roy. Soc. Lond. A} {\bf 380} (1982), no.~1778 171--185.

\bibitem{Salamon:1982}
S.~Salamon, {\it {Quaternionic K{\"a}hler manifolds}},  {\em Invent Math} {\bf 67} (1982) 143--171.

\bibitem{Mason:2007ct}
L.~J. Mason and M.~Wolf, {\it {Twistor Actions for Self-Dual Supergravities}},  {\em Commun. Math. Phys.} {\bf 288} (2009) 97--123, [\href{http://arxiv.org/abs/0706.1941}{{\tt arXiv:0706.1941}}].

\bibitem{Boyer:1985aj}
C.~P. Boyer and J.~F. Plebanski, {\it {An Infinite Hierarchy of Conservation Laws and Nonlinear Superposition Principles for Self-Dual Einstein Spaces}},  {\em J. Math. Phys.} {\bf 26} (1985) 229--234.

\bibitem{Park:1989fz}
Q.-H. Park, {\it {Extended Conformal Symmetries in Real Heavens}},  {\em Phys. Lett. B} {\bf 236} (1990) 429--432.

\bibitem{Park:1989vq}
Q.-H. Park, {\it {Selfdual Gravity as a Large $N$ Limit of the Two-dimensional Nonlinear $\sigma$ Model}},  {\em Phys. Lett. B} {\bf 238} (1990) 287--290.

\bibitem{Mason:1990}
L.~J. Mason, {\it {H-space: a universal integrable system?}},  {\em Twistor Newsletter} {\bf 30} (1990) 14--17.

\bibitem{Dunajski:2000iq}
M.~Dunajski and L.~Mason, {\it {HyperKahler hierarchies and their twistor theory}},  {\em Commun. Math. Phys.} {\bf 213} (2000) 641--672, [\href{http://arxiv.org/abs/math/0001008}{{\tt math/0001008}}].

\bibitem{Dunajski:2003gp}
M.~Dunajski and L.~J. Mason, {\it {Twistor theory of hyperKahler metrics with hidden symmetries}},  {\em J. Math. Phys.} {\bf 44} (2003) 3430--3454, [\href{http://arxiv.org/abs/math/0301171}{{\tt math/0301171}}].

\bibitem{Forgacs:1980su}
P.~Forgacs, Z.~Horvath, and L.~Palla, {\it {Towards Complete Integrability in Four-dimensions}},  {\em Phys. Rev. D} {\bf 23} (1981) 1876--1879.

\bibitem{Chau:1981gi}
L.-L. Chau, M.-l. Ge, and Y.-s. Wu, {\it {The Kac-moody Algebra in the Selfdual {Yang-Mills} Equation}},  {\em Phys. Rev. D} {\bf 25} (1982) 1086.

\bibitem{Mason:1991rf}
L.~J. Mason and N.~M.~J. Woodhouse, {\em {Integrability, selfduality, and twistor theory}}.
\newblock Oxford University Press, 1996.

\bibitem{Hitchin:1982vry}
N.~J. Hitchin, {\it {Complex manifolds and Einstein\textquoteright{}s equations}},  {\em Lect. Notes Math.} {\bf 970} (1982) 73--99.

\bibitem{Atiyah:1978wi}
M.~F. Atiyah, N.~J. Hitchin, and I.~M. Singer, {\it {Selfduality in Four-Dimensional Riemannian Geometry}},  {\em Proc. Roy. Soc. Lond. A} {\bf 362} (1978) 425--461.

\bibitem{Bu:2022iak}
W.~Bu, S.~Heuveline, and D.~Skinner, {\it {Moyal deformations, W$_{1+\infty}$ and celestial holography}},  {\em JHEP} {\bf 12} (2022) 011, [\href{http://arxiv.org/abs/2208.13750}{{\tt arXiv:2208.13750}}].

\bibitem{Alday:2022uxp}
L.~F. Alday, T.~Hansen, and J.~A. Silva, {\it {AdS Virasoro-Shapiro from dispersive sum rules}},  {\em JHEP} {\bf 10} (2022) 036, [\href{http://arxiv.org/abs/2204.07542}{{\tt arXiv:2204.07542}}].

\bibitem{Alday:2022xwz}
L.~F. Alday, T.~Hansen, and J.~A. Silva, {\it {AdS Virasoro-Shapiro from single-valued periods}},  {\em JHEP} {\bf 12} (2022) 010, [\href{http://arxiv.org/abs/2209.06223}{{\tt arXiv:2209.06223}}].

\bibitem{Alday:2023jdk}
L.~F. Alday, T.~Hansen, and J.~A. Silva, {\it {Emergent Worldsheet for the AdS Virasoro-Shapiro Amplitude}},  {\em Phys. Rev. Lett.} {\bf 131} (2023), no.~16 161603, [\href{http://arxiv.org/abs/2305.03593}{{\tt arXiv:2305.03593}}].

\bibitem{Alday:2023mvu}
L.~F. Alday and T.~Hansen, {\it {The AdS Virasoro-Shapiro amplitude}},  {\em JHEP} {\bf 10} (2023) 023, [\href{http://arxiv.org/abs/2306.12786}{{\tt arXiv:2306.12786}}].

\bibitem{Costello:2022wso}
K.~Costello and N.~M. Paquette, {\it {Celestial holography meets twisted holography: 4d amplitudes from chiral correlators}},  {\em JHEP} {\bf 10} (2022) 193, [\href{http://arxiv.org/abs/2201.02595}{{\tt arXiv:2201.02595}}].

\bibitem{Bern:1998xc}
Z.~Bern, L.~J. Dixon, M.~Perelstein, and J.~S. Rozowsky, {\it {One loop $n$ point helicity amplitudes in (selfdual) gravity}},  {\em Phys. Lett. B} {\bf 444} (1998) 273--283, [\href{http://arxiv.org/abs/hep-th/9809160}{{\tt hep-th/9809160}}].

\bibitem{Wolf:2007tx}
M.~Wolf, {\it {Self-Dual Supergravity and Twistor Theory}},  {\em Class. Quant. Grav.} {\bf 24} (2007) 6287--6328, [\href{http://arxiv.org/abs/0705.1422}{{\tt arXiv:0705.1422}}].

\bibitem{Skinner:2013xp}
D.~Skinner, {\it {Twistor strings for $ \mathcal{N} $ = 8 supergravity}},  {\em JHEP} {\bf 04} (2020) 047, [\href{http://arxiv.org/abs/1301.0868}{{\tt arXiv:1301.0868}}].

\bibitem{Freidel:2021ytz}
L.~Freidel, D.~Pranzetti, and A.-M. Raclariu, {\it {Higher spin dynamics in gravity and w1+\ensuremath{\infty} celestial symmetries}},  {\em Phys. Rev. D} {\bf 106} (2022), no.~8 086013, [\href{http://arxiv.org/abs/2112.15573}{{\tt arXiv:2112.15573}}].

\bibitem{Donnay:2024qwq}
L.~Donnay, L.~Freidel, and Y.~Herfray, {\it {Carrollian $Lw_{1+\infty}$ representation from twistor space}},  \href{http://arxiv.org/abs/2402.00688}{{\tt arXiv:2402.00688}}.

\bibitem{Hoegner:2012sq}
M.~Hoegner, {\it {Quaternion-Kaehler four-manifolds and Przanowski's function}},  {\em J. Math. Phys.} {\bf 53} (2012) 103517, [\href{http://arxiv.org/abs/1205.3977}{{\tt arXiv:1205.3977}}].

\bibitem{Lipstein:2023pih}
A.~Lipstein and S.~Nagy, {\it {Self-Dual Gravity and Color-Kinematics Duality in AdS4}},  {\em Phys. Rev. Lett.} {\bf 131} (2023), no.~8 081501, [\href{http://arxiv.org/abs/2304.07141}{{\tt arXiv:2304.07141}}].

\bibitem{Neiman:2023bkq}
Y.~Neiman, {\it {Self-dual gravity in de Sitter space: Light-cone ansatz and static-patch scattering}},  {\em Phys. Rev. D} {\bf 109} (2024), no.~2 024039, [\href{http://arxiv.org/abs/2303.17866}{{\tt arXiv:2303.17866}}].

\bibitem{Aharony:2008ug}
O.~Aharony, O.~Bergman, D.~L. Jafferis, and J.~Maldacena, {\it {N=6 superconformal Chern-Simons-matter theories, M2-branes and their gravity duals}},  {\em JHEP} {\bf 10} (2008) 091, [\href{http://arxiv.org/abs/0806.1218}{{\tt arXiv:0806.1218}}].

\bibitem{Compere:2019bua}
G.~Comp\`ere, A.~Fiorucci, and R.~Ruzziconi, {\it {The $\Lambda$-BMS$_4$ group of dS$_4$ and new boundary conditions for AdS$_4$}},  {\em Class. Quant. Grav.} {\bf 36} (2019), no.~19 195017, [\href{http://arxiv.org/abs/1905.00971}{{\tt arXiv:1905.00971}}]. [Erratum: Class.Quant.Grav. 38, 229501 (2021)].

\bibitem{Compere:2020lrt}
G.~Comp\`ere, A.~Fiorucci, and R.~Ruzziconi, {\it {The $\Lambda$-BMS$_4$ charge algebra}},  {\em JHEP} {\bf 10} (2020) 205, [\href{http://arxiv.org/abs/2004.10769}{{\tt arXiv:2004.10769}}].

\bibitem{Fiorucci:2020xto}
A.~Fiorucci and R.~Ruzziconi, {\it {Charge algebra in Al(A)dS$_{n}$ spacetimes}},  {\em JHEP} {\bf 05} (2021) 210, [\href{http://arxiv.org/abs/2011.02002}{{\tt arXiv:2011.02002}}].

\bibitem{Adamo:2023zeh}
T.~Adamo, W.~Bu, and B.~Zhu, {\it {Infrared structures of scattering on self-dual radiative backgrounds}},  \href{http://arxiv.org/abs/2309.01810}{{\tt arXiv:2309.01810}}.

\bibitem{Mago:2021wje}
J.~Mago, L.~Ren, A.~Y. Srikant, and A.~Volovich, {\it {Deformed $w_{1+\infty}$ Algebras in the Celestial CFT}},  {\em SIGMA} {\bf 19} (2023) 044, [\href{http://arxiv.org/abs/2111.11356}{{\tt arXiv:2111.11356}}].

\bibitem{Bogna:2024soon}
G.~Bogna and S.~Heuveline, {\it {Work in progress}}, .

\bibitem{Bittleston:2022nfr}
R.~Bittleston, D.~Skinner, and A.~Sharma, {\it {Quantizing the Non-linear Graviton}},  {\em Commun. Math. Phys.} {\bf 403} (2023), no.~3 1543--1609, [\href{http://arxiv.org/abs/2208.12701}{{\tt arXiv:2208.12701}}].

\bibitem{Costello:2022upu}
K.~Costello and N.~M. Paquette, {\it {Associativity of One-Loop Corrections to the Celestial Operator Product Expansion}},  {\em Phys. Rev. Lett.} {\bf 129} (2022), no.~23 231604, [\href{http://arxiv.org/abs/2204.05301}{{\tt arXiv:2204.05301}}].

\end{thebibliography}\endgroup
	\bibliographystyle{JHEP}
	
	\appendix
	
\end{document}